# On continued gravitational collapse

A. LOINGER

Dipartimento di Fisica, Università di Milano

Via Celoria, 16,  20133 Milano, Italy

**Summary.** − According to a widespread *idée fixe*, the spherically-symmetric collapse of a sufficiently massive celestial body of spherical shape should generate a black hole. I prove that this process generates simply an ordinary point mass. My argument is model-independent.

**1**. − The necessary and sufficient condition that a Riemann-Einstein space-time admit the group of spatial rotations is that its $ds^2$ be reducible to the following form [1]:

(1.1) $\qquad ds^2 = A_1(r,t)\, dt^2 - A_2(r,t)\, dr^2 - A_3(r,t)\, d\omega^2 \ , \quad (r \geq 0) \ ,$

where

(1.1′) $\qquad\qquad d\omega^2 = d\theta^2 + \sin^2\theta\, d\varphi^2 \ .$

By suitable substitutions: $r \to f_1(r,t), \ t \to f_2(r,t)$, eq. (1.1) becomes

(1.1bis) $\qquad ds^2 = B_1(r,t)\, dt^2 - B_2(r,t)\, dr^2 - r^2 d\omega^2 \ , \quad (r \geq 0) \ ,$

Let us put (cf. [2], [2bis])

(1.2) $\qquad\qquad R \equiv [r^3 + (2M)^3]^{1/3} \ , \quad (r \geq 0) \ , \ (c=G=1)$

where $M$ is the mass of a given collapsing spherical body. Accordingly, we can write

(1.3) $\qquad ds^2 = C_1(R,t)\, dt^2 - C_2(R,t)\, dR^2 - R^2 d\omega^2 \ .$

Take ideally an instantaneous photograph of our contracting sphere at any time $t = \bar{t}$; its co-ordinate radius $r_a$ be equal to $\bar{r}_a$. Then, if $\bar{R}_a \equiv [(\bar{r}_a)^3 + (2M)^3]^{1/3}$, by virtue of a well-known Birkhoff's theorem, we have (see [2], [2bis]):

$$(1.4) \qquad C_1(\bar{R}_a, \bar{t}) = \frac{\bar{R}_a - 2M}{\bar{R}_a} \quad ,$$

$$(1.4') \qquad C_2(\bar{R}_a, \bar{t}) = C_1^{-1}(\bar{R}_a, \bar{t}) \quad .$$

Now, $\bar{t}$ is just *any* time: this means that, since $r_a$ tends to zero, the star will reduce asymptotically to the origin of the space co-ordinates, i.e. it will become asymptotically a simple point mass, as described by the *original* Schwarzschild's memoirs [2], [2bis].

Remark that the original form of Schwarzschils's solution to the problem of a gravitating mass point at rest is *diffeomorphic* to the *exterior* part ($r>2M$) of the usual, standard form of solution, which is due to Hilbert [3], Droste [4], and Weyl [5]. (The invariance of the surface area $4\pi(2M)^2$ is only a geometrical curiousness, devoid of any physical significance).

**2**. − If in the mentioned standard form of solution ([3], [4], [5]) we substitute for $r$ the following function $f(r)$ (see [6]):

$$(2.1) \qquad f(r) \equiv r + 2M \quad ,$$

for the spatial region external to the collapsing body we obtain

$$(2.2) \qquad \mathrm{d}s_{ext}^2 = \frac{r}{r+2M}\mathrm{d}t^2 - \frac{r+2M}{r}\mathrm{d}r^2 - (r+2M)^2 \mathrm{d}\omega^2 \quad .$$

(Obviously, eq. (2.2) can be obtained also from Schwarzschild's $\mathrm{d}s^2$ of paper [2] with the substitution $R \rightarrow r+2M$).

In lieu of eq. (1.1bis) we have:

(2.3) $$ds^2 = D_1(r,t)\,dt^2 - D_2(r,t)\,dr^2 - (r+2M)^2 d\omega^2 \ .$$

At $t = \bar{t}$, if $r_a = \bar{r}_a$:

(2.4) $$D_1(\bar{r}_a, \bar{t}) = \frac{\bar{r}_a}{\bar{r}_a + 2M} \ ,$$

(2.4′) $$D_2(\bar{r}_a, \bar{t}) = D_1^{-1}(\bar{r}_a, \bar{t}) \ .$$

But $r_a$ tends to zero, and our object will shrink asymptotically to a point mass. Again, no black hole has been engendered by the collapsing process.

**3**. − It is commonly believed that the final stage of a collapsing rotating star is a Kerr's black hole. Now, I have proved (see [7]) that Kerr's $ds^2$ is generated in reality by a simple spinning point mass, without event horizons, stationary-limit surface, ergo-sphere. The conclusion is obvious.

Vain is the chase of the black holes.

"Nicht Jeder wandelt nur gemeine Stege:
Du siehst, die Spinnen bauen luft'ge Wege."

J. W. v. Goethe

HISTORICAL FINALE

In the Twenties of the 20th century the form of solution of the papers [3], [4], [5] was not the unique solution taken into consideration, Schwarzschild's solution [2] had not yet fallen into oblivion. (Remark that, for very good reason, only the *exterior* part, $r>2M$, of the HDW-solution was regarded as valid by all the Fathers of Relativity. Magic and science fiction were extraneous to physics; no guru had brainwashed the community of physicists).

In 1922 some witty men (*lucus a non lucendo*) proposed to make Schwarzschild's form fully equivalent to *whole* HDW-form by the assumption that Schwarzschild's $r$ take also the negative values of the interval $-2M \leq r < 0$. Obviously, the Fathers of Relativity rejected this physical and mathematical folly, which was refuted in a detailed way by Marcel Brillouin [8]. Recently, the above proposition was put forward anew by some uninformed authors, but a nonsense remains a nonsense even if it is dressed with a sauce *à la mode*.

*Acknowledgment*. − I am very grateful to my friend Dr. S. Antoci for many useful discussions and advices.